\pdfoutput=1
\documentclass[10pt]{iopart}

\usepackage{iopams}
\expandafter\let\csname equation*\endcsname\relax
\expandafter\let\csname endequation*\endcsname\relax

\usepackage{graphicx}
\usepackage{amsmath, amssymb, mathrsfs}
\usepackage{aas_macros}
\usepackage{color}
\usepackage{bm}
\usepackage[T1]{fontenc}
\usepackage[utf8]{inputenc}
\usepackage{microtype}
\usepackage[normalem]{ulem}
\usepackage[dvipsnames]{xcolor}
\usepackage[colorlinks,urlcolor=NavyBlue,citecolor=NavyBlue,linkcolor=NavyBlue,pdfusetitle]{hyperref}
\usepackage[all]{hypcap}
\usepackage{orcidlink}

\graphicspath{{analysis/plots/}}

\newcommand{\figWaveform}{%
	\begin{figure}[tb]
		\label{fig:waveform}
		\includegraphics[width=\columnwidth]{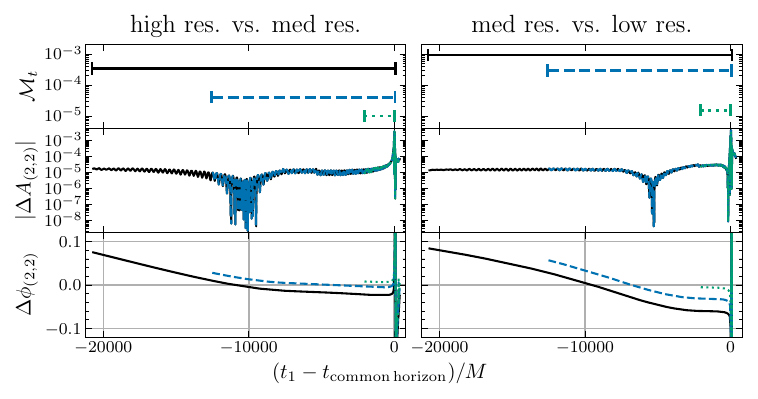}
		\caption{ \label{fig:error_components} %
			Comparison of waveforms from different resolutions for SXS:BBH:1132: a quasi-circular, equal-mass, and non-spinning simulation from the SXS catalog. The left panels are for the two highest resolutions, while the right panels are for the next two highest resolutions. The top panels show the mismatch between the waveforms when truncated at time $t_{1}$, while the middle and bottom panels show the $(2,2)$ mode's amplitude and phase error. The full simulation (black) has length $\sim$25,000$M$, and we truncate to lengths of $\sim$15,000$M$ (blue, dashed) and $\sim$2,500$M$ (green, dotted) to demonstrate the effects of simulation length. Waveforms between different resolutions are aligned to each other by optimizing over a time translation and an $SO(3)$ rotation.
		}
	\end{figure}
}

\newcommand{\figMismatch}{%
	\begin{figure}[tb]
		\label{fig:mismatch}
		\includegraphics[width=\columnwidth]{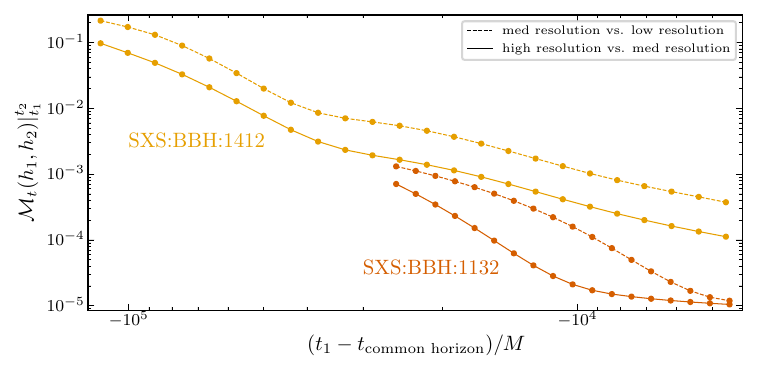}
		\caption{ \label{fig:time_dependent_mismatch} %
			Mismatch between waveforms from different resolutions for two simulations from the SXS catalog as a function of the length of the waveform. Solid lines correspond to the mismatch between the two highest resolutions, while dashed lines correspond to the mismatch between the next two highest resolutions. Waveforms between different resolutions and for different trunctations are aligned to each other by optimizing over a time translation and an $SO(3)$ rotation.
		}
	\end{figure}
}

\newcommand{\UMiss}{Department of Physics and Astronomy,
    University of Mississippi, University, MS 38677, USA}

\newcommand{\CornellPhysics}{Department of Physics, Cornell University, Ithaca,
	New York 14853, USA}

\newcommand{\CornellLepp}{Laboratory for Elementary Particle Physics, Cornell
	University, Ithaca, New York 14853, USA}

\newcommand{\CornellCcaps}{Cornell Center for Astrophysics and Planetary
	Science, Cornell University, Ithaca, New York 14853, USA}

\newcommand{\Caltech}{Theoretical Astrophysics 350-17, California Institute of
	Technology, Pasadena, CA 91125, USA}

\newcommand{\Aei}{Max Planck Institute for Gravitational Physics (Albert
	Einstein Institute), D-14467 Potsdam, Germany}

\begin{document}

\note{Length dependence of waveform mismatch:\\a caveat on waveform accuracy}

\setcounter{footnote}{0}

\author{%
	Keefe Mitman\footnotemark\,\orcidlink{0000-0003-0276-3856},
	Leo C.\ Stein$^{2}$\,\orcidlink{0000-0001-7559-9597},\\
	Michael Boyle$^{1}$\,\orcidlink{0000-0002-5075-5116},
	Nils Deppe$^{1,3,4}$\,\orcidlink{0000-0003-4557-4115},
	Lawrence E.~Kidder$^{1}$\,\orcidlink{0000-0001-5392-7342},
	Harald P.~Pfeiffer$^{5}$\,\orcidlink{0000-0001-9288-519X},
	Mark A.~Scheel$^{6}$\,\orcidlink{0000-0001-6656-9134},
}

\ead{kem343@cornell.edu}
\address{$^{1}$\CornellCcaps}
\address{$^{2}$\UMiss}
\address{$^{3}$\CornellPhysics}
\address{$^{4}$\CornellLepp}
\address{$^{5}$\Aei}
\address{$^{6}$\Caltech}

\footnotetext{Author to whom any correspondence should be addressed.}

\hypersetup{pdfauthor={Mitman, Stein, et al.},
  pdftitle={Length dependence of waveform mismatch: a caveat on waveform accuracy}}

\date{\today}

\begin{abstract}
	The Simulating eXtreme Spacetimes Collaboration's code \texttt{SpEC} can
now routinely simulate binary black hole mergers undergoing $\sim25$ orbits,
with the longest simulations undergoing nearly $\sim180$ orbits. While this
sounds impressive, the mismatch between the highest resolutions for this long
simulation is $\mathcal{O}(10^{-1})$. Meanwhile, the mismatch between
resolutions for the more typical simulations tends to be $\mathcal{O}(10^{-4})$,
despite the resolutions being similar to the long simulations'. In this note, we
explain why mismatch alone gives an incomplete picture of code---and
waveform---quality, especially in the context of providing waveform templates
for LISA and 3G detectors, which require templates with $\mathcal{O}(10^{3}) -
\mathcal{O}(10^{5})$ orbits. We argue that to ready the GW community for the
sensitivity of future detectors, numerical relativity groups must be aware of
this caveat, and also run future simulations with at least three resolutions
to properly assess waveform accuracy.
\end{abstract}

\section{Introduction}
\label{sec:introduction}

Gravitational wave (GW) science, whether it be analysis of
observed data, or quantifying waveform models' predictions, requires
an inner product on the space of waveforms.  The standard choice is a
weighted $L^{2}(\mathbb{R})$ inner product\footnote{%
  \label{fn:IR-divergence}%
  Strain waveforms with memory effects are infrared (IR) divergent and
  hence not square-integrable with the standard $L^{2}$ norm,
  i.e., with $S_{h}=1$. With detectors, this is not a problem
  since $1/S_{h} \to 0$ as $f\to 0$. Moreover, since the overlap is
  computed with some cutoff $f_{\mathrm{min}}>0$, this is not an issue.  Alternatively, one can instead use the news (the first
  time derivative of the strain) which is IR-finite.
  See Ref.~\cite{Chen:2024ieh} for improved methods of Fourier transforming
  waveforms with memory.} called the \emph{overlap}, where for two
waveforms $h_{1}$ and $h_{2}$, the overlap is
\begin{align}
	\label{eq:overlap}
	\mathcal{O}(h_{1},h_{2})=\langle h_{1}, h_{2}\rangle
  \equiv
  4\mathrm{Re}\int_{0}^{\infty}\frac{\tilde{h}_{1}(f)^{*}\ \tilde{h}_{2}(f)}{S_{h}(f)}df,
\end{align}
where tildes represent the Fourier transform, ${}^{*}$ represents
complex conjugation, and $S_{h}(f)$ is the power spectral density of
the strain noise for some GW
detector~\cite{Finn:1992xs,Cutler:1994ys,Flanagan:1997kp,Flanagan:1997sx}. This
is a natural choice for an inner product,
appearing in the Gaussian log-likelihood, and giving
the optimal\footnote{For the matched filtering SNR, see, e.g., Eqs. (109) and (110) in Ref.~\cite{Chatziioannou:2024hju}.} signal-to-noise ratio (SNR)
for a waveform $h(t)$,
\begin{align}
	\label{eq:SNR}
	\rho^{2}=||h||^{2}=\langle h, h\rangle=4\int_{0}^{\infty}\frac{|\tilde{h}(f)|^{2}}{S_{h}(f)}df.
\end{align}
When comparing waveforms, however, the overlap is often replaced by the \emph{mismatch}
\begin{align}
	\label{eq:mismatch}
	\mathcal{M}(h_{1},h_{2})=1-\frac{\langle h_{1}, h_{2}\rangle}{\sqrt{\langle h_{1}, h_{1}\rangle\langle h_{2}, h_{2}\rangle}}.
\end{align}
Note that this is often defined with an implicit optimization
over time and phase, which we do not explicitly include (see Appendix~\ref{sec:app_mismatch} for
more discussion).
Whereas overlap measures the \emph{similarity} between two waveforms,
mismatch measures the \emph{dissimilarity} of two waveforms in a
scale-invariant way~\cite{Lindblom:2008cm}. Mismatch is a useful
measure because by inserting Eq.~\eqref{eq:SNR} into
Eq.~\eqref{eq:mismatch} and expanding to leading order in $\delta h
\equiv h_{1} - h_{2}$ with $||\delta h|| \ll ||h_{1,2}||$, one finds
\begin{align}
	\label{eq:mismatchSNR}
	\mathcal{M}(h_{1},h_{2})\simeq\frac{\langle \delta h_{\perp},\delta h_{\perp}\rangle}{2\rho^{2}}\leq\frac{\langle \delta h,\delta h\rangle}{2\rho^{2}}
\end{align}
where
\begin{align}
	\delta h_{\perp}\equiv \delta h - \frac{h_{1}}{||h_{1}||}\left\langle \frac{h_{1}}{||h_{1}||}, \delta h\right\rangle.
\end{align}
Consequently, one can use the mismatch as a kind of distinguishability
criterion when comparing two waveforms or two models' predictions. That is, two
waveforms cannot be distinguished if (see App. G of Ref.~\cite{Chatziioannou:2017tdw} or
Ref.~\cite{Toubiana:2024car})
\begin{align}
	\label{eq:distinguishability}
	\mathcal{M}\lesssim\frac{D}{2\rho^{2}},
\end{align}
where $D$ is the number of dimensions that your two waveform models
depend on, e.g., 9 for binary black hole (BBH) mergers.\footnote{These 9 parameters are the mass
  ratio, two three-dimensional spin vectors, and the eccentricity and
  mean anomaly. The extrinsic parameters are not included in
  $D$, as they are model-independent.}
This criterion can also be used for testing Einstein's equations, in
which case one should study the mismatch between a beyond-GR model and
a set of approximate GR solutions.  It is crucial that the
approximate solutions' errors (e.g., analytical or numerical
truncation errors) are also smaller than this criterion, in the sense
explained in Sec.~\ref{sec:results}.

The mismatch is clearly foundational in GW science. However, it comes with many
caveats -- see App.~\ref{sec:app_mismatch} for a brief
discussion. In this paper, we highlight one issue: how mismatch
depends on the length of a numerical relativity waveform.

\section{Results}
\label{sec:results}

One area of GW science where mismatch is used extensively is numerical
relativity (NR), where the mismatch is used to determine if an NR
simulation is accurate enough to be used to analyze GW observations.
To quantify our truncation error, we run multiple NR simulations of
the same system at two or more resolutions, resulting in two
waveforms, $h_{\text{res.}\,1,2}$, and then compute
$\mathcal{M}(h_{\text{res.}\,1}, h_{\text{res.}\,2})$.  How small must
this be? As an example, suppose a detector is expected to observe BBH
mergers with SNRs of $\rho=10$; then
Eq.~\eqref{eq:distinguishability}, with $D=9$, says the truncation
error is sufficiently small when
\begin{align}
  \mathcal{M}(h_{\text{res.}\,1}, h_{\text{res.}\,2}) \lesssim 4.5\times10^{-2} \,.
\end{align}
Consequently, it is often unfortunately assumed in the NR community
that if a simulation can be run that achieves this mismatch, then the
NR code is ready to produce waveforms for the detector in question. In
reality, an NR code must achieve this mismatch \emph{over the
  detector's entire frequency range}.\footnote{%
  There is a slight caveat to this that we explain at the end of this section.}
Furthermore, as we will show now, even if one can run an NR simulation that
achieves a target mismatch, this \emph{does not} imply that the same
mismatch can be achieved for a \emph{longer} simulation.

We consider the regime where our simulations are converging
well enough that we can treat the amplitude and phase errors as small.
We write the waveform of one resolution in terms
of another,
\begin{align}
  h_{\text{res.}2}(t) = \left(1+ \frac{\delta A}{A}(t)\right) e^{i \delta\phi(t)} h_{\text{res.}1}(t) \,.
\end{align}
This parameterization is completely general, but particularly useful
when the waveforms are similar.
Here the phase error $\delta\phi(t)$ as well as the fractional amplitude
error $(\delta A/A)(t)$ are both real functions.  Amplitude errors
tend to be bounded, while phase errors are secular and steadily
accumulate as simulations run for a long time.\footnote{In the study of orbital
dynamics it is typical to use symplectic integration methods that
are able to drastically reduce such errors (often $\mathcal{O}(\Delta t^2)$
smaller than the integration method) because they preserve the Hamiltonian
dynamics exactly~\cite{1990CeMDA..50...59K, 1991AJ....102.1528W,
1993CeMDA..56...27Y, 1994Icar..108...18L, 1999MNRAS.304..793C,
2005PhRvE..71e6703S}. It is not currently known if or how to do this for
numerical relativity simulations of binary black hole mergers.}
Thus, phase errors
dominate for sufficiently long binary simulations. Now we estimate the
time-domain overlap between these two resolutions, using a flat noise
curve $S_{h}=1$, over the time interval $[t_{1},t_{2}]$. While the overlap in Eq.~\eqref{eq:overlap} is defined in the frequency domain, we stress that because of Parseval's theorem, performing this integral in the time domain is equally valid and we do so to avoid tapering complexities that are necessitated by Fourier transforms. This overlap, distinguished from its frequency-domain counterpart by a subscript $t$, is
\begin{align}
	\mathcal{O}_{t}\left(h_{\text{res.}1},h_{\text{res.}2}\right)
  = 4\mathrm{Re}\int_{t_{1}}^{t_{2}}\left(1+ \frac{\delta A}{A}\right)e^{i\delta\phi}|h_{\text{res.}1}|^{2}\,dt
    \,.
\end{align}
As the phase error is supposed to be notably small, we Taylor expand the
exponential.  Keeping in mind that we take the real part of the
integral, and everything else in the integrand is real, we must go to
second order,
\begin{align}
  \mathcal{O}_{t}\left(h_{\text{res.}1},h_{\text{res.}2}\right)
  &\approx
    4\mathrm{Re}\int_{t_{1}}^{t_{2}}
    \left[1 + \frac{\delta A}{A}(t) + i \delta\phi(t) - \frac{1}{2}(\delta\phi(t))^{2}\right]
    |h_{\text{res.}1}|^{2}\,dt
    \,.
\end{align}
For sufficiently long waveforms, phase errors dominate over amplitude
errors, and keeping only the real part yields,
\begin{align}
  \mathcal{O}_{t}\left(h_{\text{res.}1},h_{\text{res.}2}\right)
  &\approx
    4\int_{t_{1}}^{t_{2}}
    \left[1 - \frac{1}{2}(\delta\phi(t))^{2}\right]
    |h_{\text{res.}1}|^{2}\,dt
    \,,\\
  &=\mathcal{O}_{t}\left(h_{\text{res.}1},h_{\text{res.}1}\right)
    -
    2\int_{t_{1}}^{t_{2}}
    (\delta\phi(t))^{2}
    |h_{\text{res.}1}|^{2}\,dt
    \,.
\end{align}
Converting this to a mismatch, we find
\begin{align}
  \mathcal{M}_{t}\left(h_{\text{res.}1},h_{\text{res.}2}\right)
  \approx
  \frac{2}{\rho^{2}} \int_{t_{1}}^{t_{2}} (\delta\phi(t))^{2}|h_{\text{res.}1}(t)|^{2} \, dt
  \,,
\end{align}
where $\rho^{2} = \mathcal{O}_{t}\left(h_{\text{res.}1},h_{\text{res.}1}\right) =
4\int_{t_{1}}^{t_{2}} |h_{\text{res.}1}(t)|^{2} \, dt$.
The integrals in the numerator and denominator both grow with the
length of the interval $|t_{2}-t_{1}|$.  However, the numerator has
the extra factor of $(\delta\phi(t))^{2}$, making it grow faster.  As
a simple estimate, suppose we take the phase error to be a linear
function of time, $\delta\phi(t) =
(t-t_{a})(d\,\delta\phi/dt)|_{t_{a}}$, where
$t_{a}\in [t_{1},t_{2}]$ is some alignment time.  Now our mismatch is
approximated as
\begin{align}
  \label{eq:result}
  \mathcal{M}_{t}\left(h_{\text{res.}1},h_{\text{res.}2}\right)
  &\approx
  \frac{2}{\rho^{2}} \left(\frac{d \, \delta\phi}{dt}\Big|_{t_{a}}\right)^{2}
  \int_{t_{1}}^{t_{2}} (t-t_{a})^{2}|h_{\text{res.}1}(t)|^{2} \, dt\nonumber\\
  &
  \propto \left(\frac{d \, \delta\phi}{dt}\Big|_{t_{a}}\right)^{2} |t_{2}-t_{1}|^{2}
  \,,
\end{align}
where we kept the asymptotic scaling as $|t_{2}-t_{1}|$
grows very large. The integral can be performed asymptotically with, e.g., a post-Newtonian waveform.
This shows that, assuming the main error in numerical relativity simulations is a small phase error, the \emph{mismatch will tend to increase with the square of the length of the simulation.}

\figWaveform

In Fig.~\ref{fig:waveform} we first show an example illustrating that the
dominant resolution error in numerical relativity simulations tends to be the
binary phase error. In Fig.~\ref{fig:waveform} we show the $(2,2)$ mode
amplitude and phase errors between the two highest resolutions for the
Simulating eXtreme Spacetimes (SXS) Collaboration's simulation
SXS:BBH:1132~\cite{Boyle:2019kee}. We also show how these errors change as one
changes the effective length of the simulation, by removing inspiral data from
each resolution's waveform. While the amplitude error for each of the two
resolution pairs is roughly consistent and below the square root of the
mismatch,\footnote{Note that mismatch goes as the residual squared; see
Eq.~\eqref{eq:mismatchSNR}.} the phase error can be notably large. Meanwhile, in
the top panel of Fig.~\ref{fig:waveform}, we show the mismatch between the two
resolutions' waveforms. 
Here, and in subsequent plots, the mismatch is computed by integrating over the two-sphere rather than by evaluating the strain waveforms at some arbitrary point.
As one decreases the effective length of the simulation,
the mismatch correspondingly decreases, even though the amplitude and phase
errors are similar to the comparison in
which the numerical waveforms are longer. Note that here, and in subsequent figures, mismatch is computed by integrating the strain over the two-sphere.

In Fig.~\ref{fig:mismatch} we further illustrate this point regarding the length dependence of the mismatch. Each point in this figure represents taking waveforms from two resolutions, truncating them so that their length prior to the simulation's common horizon time is some $t_{1}$, minimizing the $L^{2}$ norm of their residual over the two-sphere by optimizing over a time translation and an $SO(3)$ rotation,\footnote{For an explanation of why this is done, see App.~\ref{sec:app_mismatch}.} and then computing a mismatch between $t_{1}$ and some fixed time in the late ringdown phase relative to $t_{\mathrm{common\,horizon}}$. This optimization is performed numerically using \texttt{scipy}'s \texttt{minimize} function via the \texttt{sxs} package~\cite{2020SciPy-NMeth,Boyle_The_sxs_package_2025}. Both simulations are quasi-circular, non-precessing systems. As can be seen, by making each simulation ``shorter'', one can decrease the mismatch by many orders of magnitude. Equation~\eqref{eq:result} predicts that on a log-log plot like Fig.~\ref{fig:mismatch}, we would see a slope of roughly $-2$, and indeed using \texttt{scipy}'s \texttt{curve\_fit} we find slopes in good agreement with this prediction.

At this point, one may worry that the results of Fig.~\ref{fig:mismatch} as well as the $\mathcal{M}\leq 10^{-6}$ mismatch requirement of LISA (i.e., an SNR of $\rho\sim1000$ in Eq.~\eqref{eq:distinguishability}) seem to suggest that current numerical relativity codes are no where near accurate enough to simulate the $\mathcal{O}(10^{3})$ orbit binaries that future detectors will observe. We stress, however, that this concern is partially resolved by utilizing post-Newtonian (PN) hybridizations~\cite{Sun:2024kmv}. Specifically, provided one can run a simulation long enough such that, over its length it is accurate to $\mathcal{M}\leq 10^{-6}$ and over a comparison window the NR/PN hybrid is accurate to $\mathcal{M}\leq 10^{-6}$ and PN itself before the window is accurate to $\mathcal{M}\leq 10^{-6}$,\footnote{To measure this, one unfortunately needs to compare to a long NR simulation. But performing this comparison for one point in the more complicated region of parameter space is likely sufficient to claim that PN (at whatever PN order one has) is sufficiently accurate for modeling purposes.} then one need not actually simulate $\mathcal{O}(10^{3})$ orbits. Thus far some work has been done to study how feasible this is. In particular, Ref.~\cite{Sun:2024kmv} has shown that many of the simulations produced by the SXS collaboration can obtain hybridization mismatches $\lesssim\mathcal{O}(10^{-6})$ for spin-aligned systems and $\lesssim{O}(10^{-2})$ for precessing systems, with the primary limitations being residual eccentricity and the lack of higher-order PN spin terms. Thus, with more accurate eccentricity control and higher-order PN terms, and perhaps even aid from the effective-one-body framework~\cite{Pompili:2023tna}, it is likely that PN hybridizations can ready the community for future detectors without the need to push NR to be able to accurately simulate the entirety of $\mathcal{O}(10^{3})$ orbits.\footnote{There is also a crucial need from the self-force community~\cite{Pound:2021qin} to extend the parameter space coverage provided by NR simulations to higher mass ratios, as well as an effort from all communities to understand how to model eccentric-precessing systems more generally.}

\figMismatch

\section{Discussion and conclusions}
\label{sec:disc-concl}

In this Note, we highlight a commonly overlooked feature of the
waveform mismatch: its
dependence on the length of an NR
simulation. In particular, by assuming that two waveforms differ by
some small phase error---as is the case for waveforms from different
resolution numerical relativity simulations---we showed analytically
that the mismatch tends to increase with the square of the length of
the simulation. Moreover, by examining simulations from the SXS
collaboration, we showed that this increase in mismatch can be as
large as a three orders of magnitude when increasing the length of a
simulation by $\sim 100,000M$ or $\sim 100$ orbits. Consequently,
demonstrating that a relatively short simulation is accurate to within
a target mismatch is not sufficient to claim a numerical relativity
code's accuracy for much longer simulations, i.e., those which are
required for spanning the entire frequency band of future GW
detectors.

This shows that numerical relativity groups have much to achieve
to be prepared for the next era of GW science~\cite{Purrer:2019jcp,Chandramouli:2024vhw}.
There are now many independent NR simulation
catalogs~\cite{Aylott:2009tn, Ajith:2012az, Mroue:2013xna, Hinder:2013oqa,
  Jani:2016wkt, Healy:2017psd, Healy:2019jyf, Huerta:2019oxn,
  Boyle:2019kee, Healy:2020vre, Healy:2022wdn, Ferguson:2023vta,
  Hamilton:2023qkv, Rashti:2024yoc}, but most only provide a single
resolution for each system.  We encourage future catalogs to
provide at least two, preferably three, resolutions so that their
truncation errors can be quantified.
We also point out that there are many important studies that one can pursue to
help prepare us for next-generation sensitivities: better control of
eccentricity in simulations, improving the accuracy of
PN waveforms with more terms and effective-one-body calculations, and extending mass ratio coverage with
gravitational self-force.

\ack
K.M.\ is supported by NASA through the NASA Hubble Fellowship grant \#
HST-HF2-51562.001-A awarded by the Space Telescope Science Institute,
which is operated by the Association of Universities for Research in
Astronomy, Incorporated, under NASA contract
NAS5-26555. L.C.S.\ acknowledges support from NSF CAREER Award
PHY–2047382 and a Sloan Foundation Research Fellowship. This material is based upon work supported by the National Science Foundation under Grants No.~PHY-2309211, No.~PHY-2309231, No.~OAC-2209656 at Caltech, and No.~PHY-2407742, No.~PHY-2207342, and No.~OAC-2209655 at Cornell. Any opinions, findings and conclusions or recommendations expressed in this material are those of the author(s) and do not necessarily reflect the views of the National Science Foundation. This work was supported by the Sherman Fairchild Foundation at Caltech and Cornell.

\section{References}
\bibliographystyle{iopart_num}
\bibliography{mismatch-length}

\appendix
\section{Other Mismatch Caveats}
\label{sec:app_mismatch}

The GW literature contains varying degrees of parameter minimization when
computing waveform mismatches.  Above, we only (implicitly) considered time and
$SO(3)$ rotations of two fixed waveforms.  At a different extreme, in the
context of GW detectability, one only cares if \emph{any} template has
a high overlap with a signal, so one might minimize mismatch over both
extrinsic parameters [e.g., time translations, rotations of the
source, and GW polarization angle (rotations about the line of sight
to the source)] and intrinsic parameters (e.g., masses, spins,
eccentricity), regardless of how close the template parameters are to
the true source parameters.

A more subtle question is how to use mismatch to quantify the
similarity of waveform models themselves, apart from detectors that
live at one point on the source's celestial sphere.  A waveform model
is effectively a map $f: X \to \mathcal{F}(\mathcal{I}^{+})$ where $X$ is our
parameter space (e.g., the 9-dimensional space of binary black hole mergers),
$\mathcal{I}^{+}$ is future null infinity, which topologically is
$S^{2}\times \mathbb{R}$; and $\mathcal{F}(\mathcal{I}^{+})$ is some
appropriate function space (e.g., twice differentiable for the Weyl curvature
scalars to exist; with square-integrable news for the memory to be
finite). The overlap integral can be taken over the two-sphere,
providing a metric on the function space (modulo the minor convergence
issue related to memory effects mentioned in footnote~\ref{fn:IR-divergence}).

Now we can ask how similar are two models $f,g$.  Unfortunately this
is not as simple as $\sup_{x\in X} \mathcal{M}(f(x),g(x))$ where
$\sup$ is the supremum, because the two models $f$ and $g$ may have different
meanings for the parameters; see the discussion of Fig.~1 in
Ref.~\cite{Sun:2024kmv}.  We can allow for different parameter
definitions as follows.  For some fixed waveform $f(x)$ we can find
the ``shortest distance'' from the image of $g$ by finding
$\mathcal{M}(f(x),g)=\inf_{x'\in X}\mathcal{M}(f(x),g(x'))$, where
$\inf$ is the infimum.  Doing this for all $x\in X$, we can find the
point in the image of $f$ whose shortest distance is greatest,
$\sup_{x\in X}\mathcal{M}(f(x),g)$.  Similarly, reversing the roles of
$f$ and $g$, we can find the point in the image of $g$ whose shortest
distance to the image of $f$ is greatest---this need not be the same
pair of points.  Finally we take the larger of the two mismatches.  This is the
\emph{Hausdorff distance} between the images of $f$ and $g$.
Unfortunately, this is only practical on a discrete subset
$\Lambda\subset X$, but is still a useful measure of the similarity of $f$ and $g$.

Finally, one might consider waveforms on $\mathcal{I}^{+}$ to be in
equivalence classes of future null infinity's asymptotic symmetry group, i.e., the Bondi-van der Burg-Metzner-Sachs (BMS)
group~\cite{Bondi1960,Sachs1961,Bondi:1962px,Sachs1962PR,Sachs:1962wk,Mitman:2024uss}. Some of these BMS transformations are clearly physically
meaningful across the set of all binary merger observations---for
example, each merger happens at a specific time, so it can not
be translated to a different time.  However in the model, we have a
universe with only a single GW source.  The action of BMS on the
function space $\mathcal{F}(\mathcal{I}^{+})$ gives gauge-equivalent
orbits, so a measurement of model similarity should be a
Hausdorff distance on the quotient space
$\mathcal{F}(\mathcal{I}^{+})/\text{BMS}$. Note that this is partially achieved when performing the optimizations described earlier, i.e., optimizing over time translations, source rotations, and GW polarization angle, but one should also include the Lorentz boost and supertranslation transformations.

\end{document}